\documentclass[twoside]{dis04} \usepackage{amssymb}
\def\runauthor{``T.C. Rogers and M.I. Strikman"} \def\shorttitle{The
Black Body Limit in DIS}
\usepackage{amsmath}
\begin{document}

\title{The Black Body Limit in Deep Inelastic Scattering \\ }

\author{T.C. Rogers and M.I. Strikman}

\address{Pennsylvania State University\\ University Park, Pennsylvania
16802, U.S.A.\\ E-mail: rogers@phys.psu.edu }

\maketitle

\abstracts{We use information from DIS and the two gluon nucleon form
factor to estimate the impact parameter  amplitude of hadronic
configurations in the dipole model of  DIS.  We demonstrate that only a
small fraction of the total  $\gamma^{\ast}N$ cross section at $x\sim
10^{-4}$  is due to scattering that occurs near the black body limit.
We also make comparisons with other models and we point out that 
a quark mass of $\lesssim 100$~MeV leads to a strong variation 
of the the $t$-dependence with $Q^{2}$.  \\ }

\section{Introduction} 
When one assumes that the scattering amplitude for hadron-hadron
scattering is purely imaginary, unitarity implies that the total
elastic cross section, $\sigma_{el}(b)$, is  less than the total
inelastic cross section, $\sigma_{in}(b)$,  at impact parameter, $b$
(a small real part leads to a small correction).  The unitarity limit
is  saturated when $\sigma_{el}(b) = \sigma_{in}(b)$, which is
equivalent to the condition that the profile function defined by,
$\Gamma_{h}(s,b) \equiv \frac{1}{2is(2 \pi)^{2}} \int d^{2}  \vec{q}
e^{i \vec{q} \cdot \vec{b}} A_{hN}(s,t)$, is equal to unity.  When
this situation is reached, the target is totally absorbing at impact
parameter, b, and scattering  is said to occur in the black body limit
(BBL).  See, for example, Ref.~\cite{Gribov:1968ia}.

Within the dipole model of DIS, the total $\gamma^{\ast}N$  cross
section is written as the convolution of the basic cross section for
the interaction of the virtual photon's  hadronic constituents with
the square modulus of the photon  wave function:

\begin{eqnarray}
\sigma^{ \gamma^{\ast} N}_{L,T}(Q^{2},x) = \int_{0}^{1}  dz \int d^{2}
\vec{d}   \left| \psi_{L,T} (z,d) \right|^{2}
\hat{\sigma}_{tot}(d,x^{\prime}), \label{eq:totalcross}
\end{eqnarray}
where $x^{\prime}$ is a function of $Q^{2}$.  We will discuss the nearness of
the hadronic constituent  cross sections,
$\hat{\sigma}_{tot}(d,x^{\prime})$, to the BBL.  We use the model of 
McDermott, Frankfurt, Guzey, and Strikman (MFGS),
discussed in Ref.~\cite{McDermott:1999fa}, to  interpolate between the
hard and soft regimes for the  total cross section.  For the
t-dependence, discussed in  section~2, we use a model discussed in
Ref.~\cite{Rogers:2003vi}.  In the  next section, we will demonstrate
that a non-negligible fraction of the total cross section receives
contributions from hadronic scattering near the BBL.   In section~3,
we will compare our model of the basic  hadronic cross section with
other models, and we will discuss the necessity of using constituent
quark masses in the photon wave function.

\section{Modeling the t-dependence}
\begin{figure}
\centering
\rotatebox{270}{{\includegraphics[scale=0.30]{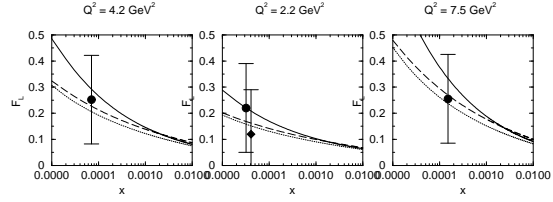}}}
\caption[*]{\label{fig:data}MFGS model prediction for $F_L$ vs. $x$
showing consistency with preliminary HERA data.  The circular points
correspond to H1 mb99(shape) and the diamond shaped point corresponds
to H1 svtx00(shape) (see Ref.~\cite{Lobodzinska}).  The dashed,
solid and dotted curves correspond to the different parton
distributions, CTEQ6L, CTEQ5L, and MRST98
respectively~\cite{Lai:1999wy,Pumplin:2002vw,Martin:1998sq}.}
\end{figure}

\begin{figure}
\centering
\rotatebox{270}{{\includegraphics[scale=0.30]{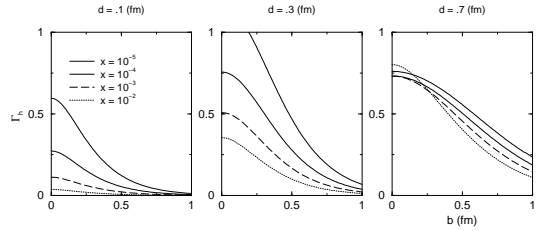}}}
\caption[*]{\label{fig:figure1}Samples of the profile function for
different hadronic sizes.}
\end{figure}

\begin{figure}
\centering
\rotatebox{270}{{\includegraphics[scale=0.30]{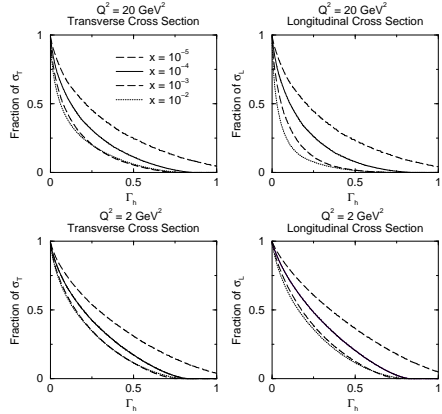}}}
\caption[*]{\label{fig:figure2}The fraction of the total cross section
due to large values of the profile function.  Note the small
difference between the longitudinal and transverse cross sections.  We
believe this to be a numerical effect.}
\end{figure}

We use the following model for the basic hadronic amplitude:
    
\begin{eqnarray}
A_{hN}(s,t) = i s \hat{\sigma}_{tot} \frac{1}{ (1  -
  t/M^{2}(d^{2}))^{2}} \frac{1}{1 - t d^{2} /d_{\pi}^{2} m_{2}^{2}}
  e^{ \alpha^{\prime} \frac{d^{2} t}{d_{\pi}^{2}}  \ln \frac{x_{0}}{x}}
\label{eq:ourmodel},
\end{eqnarray}
with,
\begin{eqnarray}
M^{2}(d^{2}) = \left\{ \begin{array}{ll} m_{1}^{2} - (m_{1}^{2} -
             m_{0}^{2}) \frac{d^{2}}{d_{\pi}^{2}}  &
             ,\mbox{\hspace{2mm}$d \leq d_{\pi}$} \\
                                                               
	    m_{0}^{2} & ,\mbox{\hspace{2mm}otherwise}
                       \end{array} \right. \,. \label{eq:M}
\end{eqnarray}
(See Ref.~\cite{Rogers:2003vi} for the detailed procedure.)   Here,
$m_{1}^{2} \approx 1.1$~GeV$^2$, $m_{0}^{2}  \approx 0.7$~GeV$^2$ and
$m_{2}^{2} \approx 0.6$~GeV$^2$.  The  typical size of the pion is
$d_{\pi} \approx 0.65$~fm,  $\alpha^{\prime}$ is $0.25$, and $x_{0} =
.01$.  It can be checked that this form of the amplitude reproduces
the expected small size, hard behavior while yielding the
phenomenologically known behavior of the larger size, soft
configurations.  The exponential factor in Eq.~\ref{eq:ourmodel} takes
into account scattering by soft Pomeron exchange.
Figure~\ref{fig:figure1} shows some samples of the resulting profile
function for small and larg hadronic sizes, and
Fig.~\ref{fig:figure2} demonstrates that a small fraction of the total
cross section, obtained from Eq.~\ref{eq:totalcross}, is due to
scattering near the BBL.  We would like to stress that the MFGS model
makes reasonable predictions for $F_L$ at low $Q^{2}$ which is
demonstrated in Fig.~\ref{fig:data} which makes comparisons between
the MFGS  predictions and preliminary data from
Ref.~\cite{Lobodzinska}.  This parameterization is consistent with
HERA data on $J/ \Psi$~\cite{Frankfurt:2000ez}.

\begin{figure}
\centering
\rotatebox{270}{{\includegraphics[scale=0.25]{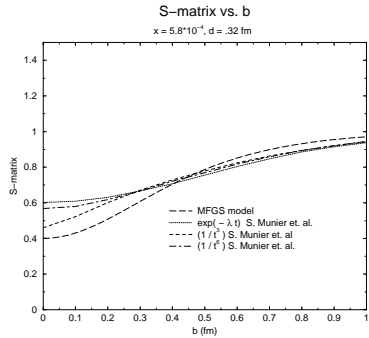}}}
\caption[*]{\label{fig:figure3}Consistency of the MFGS model with one
using $\rho$ meson production alone.  The dotted, dashed and
dot-dashed curves are taken from Ref. \cite{Munier:2001nr}.}
\end{figure}

\begin{figure}
\centering
\rotatebox{270}{{\includegraphics[scale=0.25]{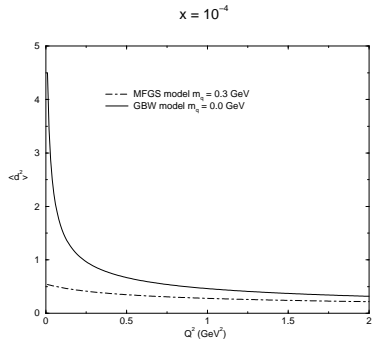}}}
\caption[*]{\label{fig:figure4}Average squared hadronic size for the
MFGS model and the GBW model.  This plot demonstrates the  significant
sensitivity to the quark mass at low $Q^{2}$.}
\end{figure}

\begin{figure}
\centering
\rotatebox{270}{{\includegraphics[scale=0.30]{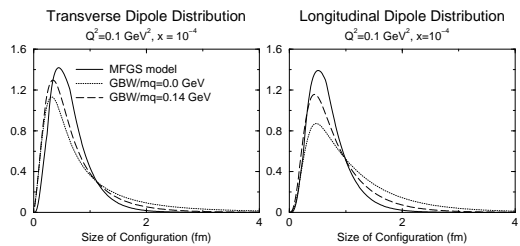}}}
\caption[*]{\label{fig:figure5}Normalized distribution of hadronic
sizes for our model with light quark masses equal to $0.3$~GeV$^2$ and
charm mass equal to $1.5$~GeV$^2$, compared with the saturation model
with all quark masses set to zero (dotted line) and with all masses
equal to $0.14$~GeV$^2$ (dashed line).}
\end{figure}

\section{Comparisons with other models}

A model similar to ours, but which only uses $\rho$-meson production
is  discussed in Ref.~\cite{Munier:2001nr}.  This model yields
qualitatively similar results to what we have found~\cite{thank}.  A
comparison between these two models is shown in
Fig.~\ref{fig:figure3}.  (Note the definition, $S(b) = 1 -
\Gamma(b)$.)  We would like to  stress that data for $\rho$ production
is limited to kinematics where  $-t \lesssim 0.6$~GeV$^2$ ($b \gtrsim
0.3$~fm), whereas the MFGS model is valid down to much smaller values
of $-t$.

The model of Golec-Biernat and Wusthoff
(GBW)~\cite{Golec-Biernat:1998js,Golec-Biernat:1999qd}, and its
extension in Ref.~\cite{Bartels:2002cj}, imposes exponential taming at
large hadronic sizes  (soft physics).  In order to match the large
size  behavior to the perturbative, small size regime, the GBW model
requires that the light quark masses remain small ($0.0$ to
$0.14$~GeV) even in the large size,  nonperturbative regime.  In the
MFGS model, the light quark mass in the soft  regime is fixed at a
value of the order of hadronic masses (i.e., constituent quark  mass).
Form factors in the hard regime are naturally nearly independent of
the quark mass.  However, at small $Q^{2}$ the cross section becomes
very sensitive to the light quark mass.  This is demonstrated  in
Fig.~\ref{fig:figure4} which shows large variation of the average
hadronic size between the MFGS model and the GBW model.  The small
quark  mass which is forced upon the model of Golec-Biernat and
Wusthoff probably overestimates the contribution from large size
configurations at small $Q^{2}$.  Figure~\ref{fig:figure5}
demonstrates the large contribution from large size configurations in
the model of Golec-Biernat  and Wusthoff for $Q^{2}=0.1$~GeV$^2$.
Thus, we conclude that, for an exclusive process  like $\gamma^{\ast}
p \rightarrow \gamma p$, and for $0 \lesssim Q^{2} \lesssim
.1$~GeV$^2$, the $t$-dependence in the GBW model varies strongly with
$Q^2$.

\section{Conclusion}
We conclude that effects from the proximity to the BBL will be small
for $x \lesssim 10^{-4}$ for $Q^{2} \approx 2$~GeV$^2$.  With the
extra color factor of 9/4 that appears when we consider a gluon
dipole, the interaction is closer to the BBL.  This is consistent with
diffractive HERA data.  We also note that our model has qualitative
consistency with other dipole models based on different methods.  We
have stressed that special care must be taken in choosing the quark
masses at small values of $Q^{2}$.  We also wish to point out that
lattice QCD calculations and approximations from instanton models
support the use of large light quark masses in the small $Q^{2}$
regime.  (See Ref.~\cite{Diakonov:2002fq} for an overview of
nonperturbative calculations.)

\section*{Acknowledgements}
We would like to thank J. Collins, L. Frankfurt, A. Mueller,
C. Weiss, M. Diehl, and X. Zu for useful discussions.

\end{document}